# Supercurrent reversal in quantum dots

Jorden A. van Dam[1], Yuli V. Nazarov[1], Erik P. A. M. Bakkers[2], Silvano De Franceschi[1,3], & Leo P. Kouwenhoven[1]

[1]Kavli Institute of Nanoscience, Delft University of Technology, PO Box 5046, 2600 GA, Delft, The Netherlands. [2]Philips Research Laboratories, Professor Holstlaan 4, 5656 AA Eindhoven, The Netherlands. [3]Laboratorio Nazionale TASC-INFM-CNR, Area Science Park, S.S. 14, Km. 163.5, I-34012 Trieste, Italy.

**When two superconductors become electrically connected by a weak link a zero-resistance supercurrent can flow[1,2]. This supercurrent is carried by Cooper pairs of electrons with a combined charge of twice the elementary charge, *e*. The *2e* charge quantum is clearly visible in the height of Shapiro steps in Josephson junctions under microwave irradiation and in the magnetic flux periodicity of *h/2e* in superconducting quantum interference devices[2]. Several different materials have been used to weakly couple superconductors, such as tunnel barriers, normal metals, or semiconductors[2]. Here, we study supercurrents through a quantum dot created in a semiconductor nanowire by local electrostatic gating. Due to strong Coulomb interaction, electrons only tunnel one-by-one through the discrete energy levels of the quantum dot. This nevertheless can yield a supercurrent when subsequent tunnel events are coherent[3-7]. These quantum coherent tunnelling processes can result in either a positive or a negative supercurrent, i.e. in a normal or a π-junction[8-10], respectively. We demonstrate that the supercurrent reverses sign by adding a single electron spin to the quantum dot. When excited states of the quantum dot are involved in transport, the supercurrent sign also depends on the character of the orbital wavefunctions.**

The electronic properties of quantum dots can be probed by attaching a source and drain electrode, allowing charge carriers to tunnel from the dot to both electrodes. If the electrodes are superconducting, transport is strongly affected and largely depends on the transparency of the electrical connection between the electrodes and the quantum dot. A number of experiments have focused on various phenomena in the Coulomb blockade regime but no supercurrents through quantum dots were observed, mostly due to the lack of a controllable tunnel coupling with the electrodes[11-14]. Strong coupling and negligible Coulomb interactions were recently obtained in carbon nanotube quantum dots demonstrating resonant tunnelling of Cooper pairs through a single quantum state[15]. In the regime of strong Coulomb interactions the simultaneous occupation of the quantum dot with two electrons is unfavourable. Nevertheless a supercurrent can flow due to the subsequent (but coherent) transport of correlated electrons. This can give rise to a sign change of the Cooper pair singlet (i.e. from $(|\uparrow\downarrow\rangle-|\downarrow\uparrow\rangle)/\sqrt{2}$ to $e^{i\pi}(|\uparrow\downarrow\rangle-|\downarrow\uparrow\rangle)/\sqrt{2}$). Therefore, the typical Josephson relation between the supercurrent, $I_s$, and the macroscopic phase difference between the superconductors, $\varphi$, usually given by $I_s=I_c \cdot sin(\varphi)$, changes to $I_s=I_c \cdot sin(\varphi+\pi)=-I_c \cdot sin(\varphi)$ (ref. 5, $I_c$ is the critical current). Other mechanisms of Cooper pair transport resulting in negative supercurrents have been studied using high-$T_c$ superconductors[8], ferromagnets[9], and non-equilibrium mesoscopic normal metals[10].

We use indium arsenide (InAs) nanowires as semiconductor weak links[16] in combination with local gate electrodes in order to obtain quantum dots with a tunable coupling to superconducting leads. The mono-crystalline n-type InAs nanowires are grown by a catalytic process based on the vapour-liquid-solid growth method[17-20]. After growth, the wires are transferred to an oxidized silicon substrate. Previously developed nanofabrication techniques are used to define highly-transparent aluminium-based superconducting contacts[16]. Pairs of nearby nanowires are contacted in parallel forming a superconducting loop with two nanowire junctions (Fig. 1a). In a second lithographic step, we define local gate electrodes. One of the nanowires (top nanowire in Fig. 1a) is crossed by two gates, labelled *L* and *R*, in order to define a quantum dot (also see Fig. 1b). The bottom nanowire is crossed by one gate, labelled *REF*, and will be used as a reference junction with a tunable Josephson coupling. We have studied two similar devices in detail. Here we present the results for one of them. Similar data from the second device and further details on device fabrication are given as Supplementary Information.

Below the superconducting transition temperature of the aluminium-based contacts ($T_c \approx 1.1$K), the two nanowires form superconducting weak links due to the proximity effect[16], thereby realizing a quantum interference device (SQUID)[2]. The critical current of the SQUID, $I_c$, as a function of magnetic flux, $\Phi$, shows oscillations with a period of 66μT. This is consistent with the addition of a flux quantum, $\Phi_0=h/2e$ (*h* is Planck's constant, *e* the electron charge), to the effective SQUID area of 30μm$^2$ (Fig. 1c, blue trace, *T*=30mK). The maximum (minimum) critical current corresponds to the sum (difference) of the critical currents of the two nanowire junctions. Unlike in other SQUIDs, the critical currents of the individual junctions can be tuned by applying voltages to the respective gates. This is demonstrated by a measurement of the SQUID oscillations for different voltages applied to *REF*. When $V_{REF}$=-0.64V (green trace) the amplitude of the SQUID oscillations is reduced due to the partial local depletion of the nanowire. By further reducing the gate voltage to $V_{REF}$=-0.80V, the reference junction is pinched off resulting in the disappearance of the interference signal. We thus have a unique electrical control over the SQUID operation.

A quantum dot is formed in the top nanowire by applying negative voltages simultaneously to gates *L* and *R*. The local depletion creates two tunnel barriers which define a single quantum dot in the nanowire section between the gates (see inset to Fig. 2a) giving rise to discrete energy levels and Coulomb blockade. To show this we pinch off the reference junction ($V_{REF}$=-0.80V) and apply a small magnetic field in order to suppress superconductivity. Figure 1d shows a colour plot of absolute current through the quantum dot, |*I*|, as a function of bias voltage, *V*, and gate voltages, $V_L=V_R$. Coulomb blockade (|*I*|=0) occurs within continuous diamond-shaped regions as it is typically observed in transport through single quantum dots[21]. Outside these regions, |*I*| increases in steps (lines parallel to the diamond edges) denoting the onset of single-electron tunnelling via discrete excited states. From the separation between these lines we estimate for this regime a characteristic level spacing of ~1 meV. The sharpness of the diamond edges and the excitation lines denote a weak tunnel coupling between the quantum dot and the source and drain leads. We can increase the coupling by reducing the negative voltages applied to *L* and *R* (see Fig. 1e). This results in smoother diamond edges (dotted lines) and the appearance of inelastic co-tunnelling features inside the diamonds. This tunable coupling is particularly important for reaching the narrow transport regime where charging effects dominate but, at the same time, the critical current is large enough to be measurable.

Switching to the superconducting state but with the reference junction still pinched off, two peaks in $dI/dV$ develop around $V \approx \pm 200\mu V = \pm 2\Delta^*/e$ (Fig. 2b). $2\Delta^*$ is the superconducting gap induced in the nanowire by the proximity effect (inset Fig. 2a). These features are due to second-order co-tunnelling, and the peak shape reflects the singularities in the quasi-particle density of states at the gap edges. In spite of the Coulomb blockade effect, we observe a finite supercurrent, $I_{c,qd}$, through the nanowire quantum dot. We exploit the SQUID geometry to determine the critical value and the sign of this supercurrent in a current-biased measurement[22]. When an integer number of flux quanta is applied through the SQUID area the critical current of the SQUID corresponds to the sum of the critical currents of the two junctions[2], i.e. $I_c = I_{c,qd} + I_{c,REF}$. We set $I_{c,REF}=320$pA, and extract the $V_L$-dependence of $I_{c,qd}$ directly from the measurement of $I_c$ (Fig. 2a). We find $I_{c,qd}<0$ for two charge states of the quantum dot, denoted by ◊, and □ in Fig. 2b. The negative supercurrent of the quantum dot junction is confirmed by the $\Phi_0/2$-shift between the SQUID oscillations for $I_{c,qd}<0$ (Fig. 2c, red trace) and those for $I_{c,qd}>0$ (blue trace). A colour plot of $I_c(V_L,\Phi)$ in Fig. 2d shows the transitions between positive and negative supercurrents around the charge state denoted by □.

Negative supercurrents have been predicted for superconductors coupled by a magnetic impurity or a single-level interacting quantum dot[3-5,7]. In these systems resonant tunnelling of Cooper pairs is prohibited due to Coulomb blockade. Nevertheless, Cooper pairs can be transported via fourth-order co-tunnelling events. Three examples of such events are shown in Fig. 3. The top and bottom diagrams are the initial and final states, respectively, and the diagrams in between show one of the three intermediate virtual states. Due to Coulomb blockade, a sequence of intermediate states involves an energy cost comparable to the charging energy, $E_c$ (for $\Delta^* \ll E_c$). Nevertheless, when the tunnel rate is on the order of $E_c/h$, a Cooper pair can be transported by higher order co-tunnelling events[23]. In principle, there are 24 possible sequences of 4 tunnel events. However, in a single-level quantum dot only a small number of sequences are allowed. Figure 3a illustrates the transfer of a Cooper pair through a quantum dot with a single spin-degenerate level occupied by one electron (with spin up, $|\uparrow\rangle$). The sequence of four tunnel processes, indicated by the numbers, is necessarily permuted compared to ordinary transport of Cooper pairs. The remarkable result is that the spin-ordering of the Cooper pair is reversed, that is, the Cooper pair on the right is created in the order $|\uparrow\rangle$, $|\downarrow\rangle$ while the pair on the left is annihilated in the order $|\downarrow\rangle$, $|\uparrow\rangle$. This spin-reversal results in a sign-change of the Cooper pair singlet state (e.g. from $(|\uparrow\downarrow\rangle-|\downarrow\uparrow\rangle)/\sqrt{2}$ to $e^{i\pi}(|\uparrow\downarrow\rangle-|\downarrow\uparrow\rangle)/\sqrt{2}$) leading to a $\pi$-shift in the Josephson relation and a negative supercurrent. However, if an extra electron is added to the quantum dot the sequence of tunnel events discussed above is prohibited due to the Pauli exclusion principle. Now other sequences of tunnel events are allowed which result in a normal, positive supercurrent[7] (see Fig. 3b). Therefore, in a single-level quantum dot a negative (positive) supercurrent is expected for an odd (even) number of electrons.

We can discriminate between odd and even numbers of electrons in Fig. 2b by measuring the linear conductance, $G$, as a function of gate voltage and magnetic field, $B$ (see Fig. 2e). We observe that the Coulomb peak spacing for the two charge states denoted by □ and ◊ increases due to the Zeeman effect, demonstrating that for these charge states the occupation number, $n$, is odd[21] (only □ is shown, |g-factor|≈15 similar to previous results for similar systems[24]). These observations are consistent with the model described above.

However, for the charge state around $V_L$=-447mV with an odd number of electrons we observe a very small, but positive critical current ($I_c$≈10pA). Moreover, in a different gate voltage range, shown in Fig. 4a, supercurrent reversal is observed also for charge states with an even number of electrons. We argue that these observations originate from co-tunnelling via multiple energy levels of the quantum dot. The multi-level nature of the quantum dot for the gate range studied in Fig. 4a emerges from the measurement of differential conductance in the normal-state (Fig. 1e). Here several peaks parallel to the diamond edges are observed, which correspond to transport through excited states of the quantum dot. In this gate voltage range the level spacing, $\delta$, is of the order of $E_c$. Therefore, these excited states can take part in co-tunnelling events and the simple model of a single-level quantum dot is no longer appropriate. As a result, all 24 sequences of tunnel events are allowed for both odd and even numbers of electrons. Therefore, a negative supercurrent due to permutation of tunnel events is possible for all values of $n$[7,25].

Additionally, in the multi-level regime properties of the wavefunctions of the quantum dot become important. To illustrate this we consider the co-tunnelling event in Fig. 3c in which two different energy levels are involved in a dot with an even number of electrons. Because the two electrons take a different path they can acquire a different phase. The opposite parity of the wavefunctions results in a phase difference of $\pi$ and therefore this event contributes to a negative supercurrent[7] (see Supplementary Information). So, for a multi-level dot two effects can result in supercurrent reversal: permutation of tunnel events and an opposite parity of wavefunctions. When co-tunnelling events with a negative contribution dominate, the junction will exhibit a negative supercurrent. We note that the presence of the Kondo effect[26,27] can result in a positive supercurrent where otherwise a negative supercurrent would be expected[28,29]. We have not found any evidence for the Kondo effect in the normal state and therefore we disregard Kondo correlations in the modeling.

To further investigate the importance of multi-level effects we numerically evaluate the critical current using fourth-order perturbation theory[6,7] (see Supplementary Information for details). We assume that tunnel couplings are random in amplitude and sign (reflecting the parity of wavefunctions) and set $\Delta^*/E_c$=0.1, as in our experiment. The probability for a negative supercurrent in the centre of a Coulomb diamond, $P_\pi$, is plotted in Fig. 4b for odd and even numbers of electrons. A very large average level spacing ($\delta/E_c$»1) effectively gives a single-level quantum dot so that $P_\pi$=1 (0) for odd (even) numbers as explained in Figs. 3a,b. The dependence of the critical current, $I_{c,qd}$, on $V_{gate}$ (Fig. 4c) indeed unambiguously demonstrates the correlation between the number of electrons on the dot and the supercurrent sign. This correlation is absent in the opposite limit ($\delta/E_c$«1), where $P_\pi$≈0.3 for both odd and even numbers of electrons, in agreement with previous calculations[7]. From the experimental data in Fig. 1e, we estimate $\delta/E_c \approx 0.4$ which clearly indicates an intermediate regime. Fig. 4d shows a typical result for $I_{c,qd}$ versus $V_{gate}$ for $\delta/E_c=0.4$. As observed in the experiment we obtain a negative supercurrent for both even (blue dot) and odd (red dot) numbers of electrons. Also the typical line-shapes are in close resemblance with the experimental data. Thus, in this multi-level regime co-tunnelling events occur through a single level as well as through different levels. Consequently, the sign of the supercurrent is not only determined by the number of electrons on the quantum dot but also by the wavefunctions of the energy levels.

We thank Y-J. Doh and L. Glazman for discussions, G. Immink for nanowire growth and A. van der Enden and R. Schouten for technical support. Financial support was obtained from the Dutch Foundation for Fundamental Research on Matter (FOM), the Dutch Organisation for Scientific Research (NOW), the EU programs HYSWITCH and NODE, and the Japanese International Cooperative Research Project (ICORP).

**Figure 1.** Sample layout and device characterization. **a,** Scanning electron micrograph of the InAs nanowire SQUID. Two nanowires (diameter≈60nm) are incorporated in a superconducting loop (100 nm Al on 10 nm Ti). Aluminium top-gates (*L* and *R*) with a spacing of ~65 nm are used to define a quantum dot in the top nanowire. A third gate (*REF*) is used to control the reference junction. **b,** High-resolution image of the top nanowire shown in (a). **c,** Critical current of the SQUID, $I_c$, versus magnetic flux, $\Phi$, for different voltages applied to the reference gate ($V_{REF}$=0V (blue), -0.64V (green), and -0.80V (red)) demonstrating full electrical control over the amplitude of the SQUID oscillations. **d,** Colour plot of absolute current through the dot, |*I*|, (increasing from white (0 pA) to red (5 pA)) versus source-drain bias voltage, *V*, and $V_L$=$V_R$ in the normal state. The Coulomb diamonds are well defined due to the weak tunnel coupling between quantum dot and leads. **e,** Differential conductance, *dI/dV*, (increasing from white (0.1µS) to red (40µS)) as a function of *V* and $V_L$ ($V_R$=-0.40V). The stronger dot-lead coupling results in blurred diamond edges (indicated by dotted lines) and horizontal features inside the diamonds due to inelastic co-tunnelling. Data in (d) and (e) are taken at *T*=30mK, and in a small magnetic field to drive the superconducting contacts into the normal state.

**Figure 2.** Supercurrent reversal in an interacting quantum dot. **a,** Plot of the critical current of the quantum dot, $I_{c,qd}$, as a function of gate voltage, $V_L$, for the same gate voltage region as in (b). A negative supercurrent is observed for two charge states. Inset: schematic of the quantum dot in the nanowire. **b,** Colour scale plot of differential conductance, $dI/dV(V,V_L)$, in the superconducting state (*dI/dV* increases from blue, white, to red. $V_{REF}$=-0.8V). The two peaks in *dI/dV* at *V*≈±200µV are due to quasiparticle co-tunnelling and their spacing (4Δ* as indicated) provides a direct measurement of the induced superconducting gap in the nanowire. □, ◊ indicate two charge states that exhibit negative supercurrent. Blue dotted lines indicate the diamond edges. **c,** Two $I_c(\Phi)$ curves taken at gate voltages indicated by the vertical red and blue dotted lines in (b), demonstrating the shift by $\Phi_0/2$ between the conventional (blue) and the π-regime (red). **d**, Critical current of the SQUID, $I_c$, in colour-scale as a function of magnetic flux, $\Phi$, and gate voltage, $V_L$. □: The interference signal is shifted by half a flux quantum compared to adjacent Coulomb diamonds, indicating the π-shift in the Josephson relation. Red and blue dotted lines correspond to red and blue traces in (c). Individual $I_c(\Phi)$ curves can be fitted very well with a sine-function within the measurement accuracy. **e,** Gray-scale plot of linear conductance, *G* (increasing from black to white), as a function of magnetic field, *B*, and gate voltage, $V_L$. □: The Coulomb peak spacing in this charge state increases with increasing field due to the Zeeman effect, indicating that the number of electrons is odd. All measurements are taken at *T*=30mK. Note that the measurements in (a), (c), and (d) are current-biased and in (b) and (e) voltage-biased.

**Figure 3.** Energy diagrams illustrating Cooper pair transport through a quantum dot due to fourth-order co-tunnelling. Top and bottom panels represent initial and final states, respectively. The intermediate panels show one of the three virtual intermediate states. Numbers indicate the sequence of tunnel events. Red (blue) corresponds to the tunnelling of a spin-down (spin-up) electron. **a,**

Transport occurs through a single spin-degenerate level filled with one electron. During this event the spin-ordering of the Cooper pair is reversed. This results in a negative contribution to the supercurrent (see also diagrams in ref. 5). **b**, Transport through one spin-degenerate level filled with two electrons. The spin-ordering of the Cooper pair cannot be reversed, resulting always in a positive supercurrent. **c**, Co-tunnelling event involving two energy levels with wavefunctions of opposite parity. This results in a negative contribution to the supercurrent[7].

**Figure 4.** Experimental results and numerical simulations for a multi-level quantum dot. Panels are ordered clockwise. **a**, Measured critical current of the quantum dot, $I_{c,qd}$, as a function of $V_L$ showing supercurrent reversal for even and odd numbers of electrons (indicated by a blue and red dot, respectively, $V_R$=-0.4V). **b**, Calculated probability of $\pi$-behaviour, $P_\pi$, for odd (red) and even (blue) numbers of electrons as a function of $\delta/E_c$. Strength and sign of tunnel couplings are randomly varied. For $\delta/E_c \gg 1$, the limiting case of a single-level quantum dot is reached, resulting in $\pi$-behaviour for odd numbers and conventional behaviour for even numbers of electrons. In the multi-level limit ($\delta/E_c \ll 1$) we obtain $P_\pi \approx 0.3$ for both even and odd numbers of electrons. **c,** Calculated critical current, $I_{c,qd}$, as a function of gate voltage, $V_{gate}$, for $\delta/E_c$=2. For odd numbers of electrons (red dots) the critical current is typically negative, similar to the measurement shown in Fig. 2a. **d**, $I_{c,qd}(V_{gate})$ for $\delta/E_c$=0.4. Negative supercurrents are found for both odd (red dot) and even numbers of electrons (blue dot) like in the experimental data shown in (a).

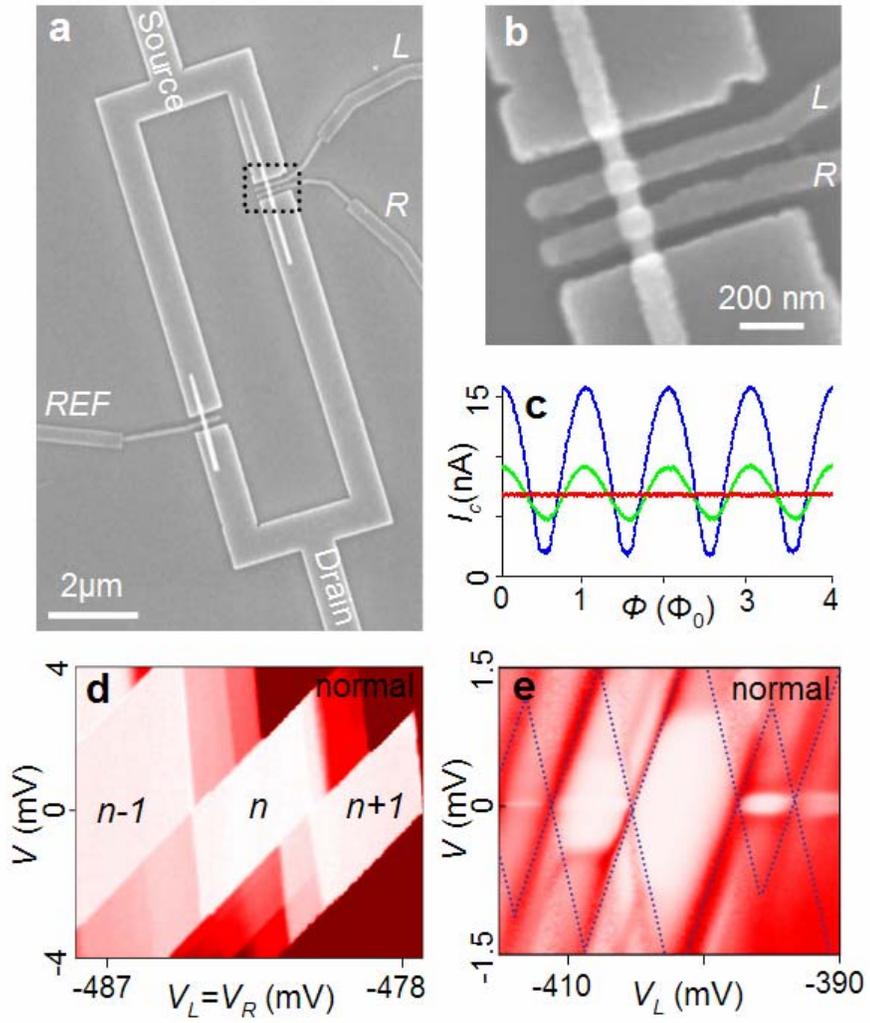

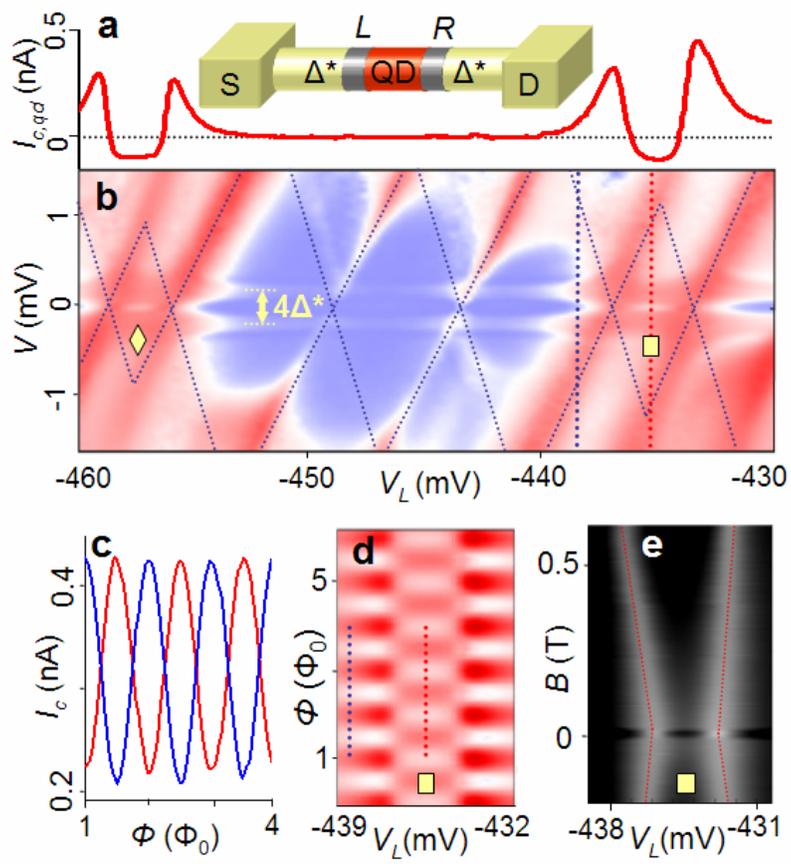

Figure 3

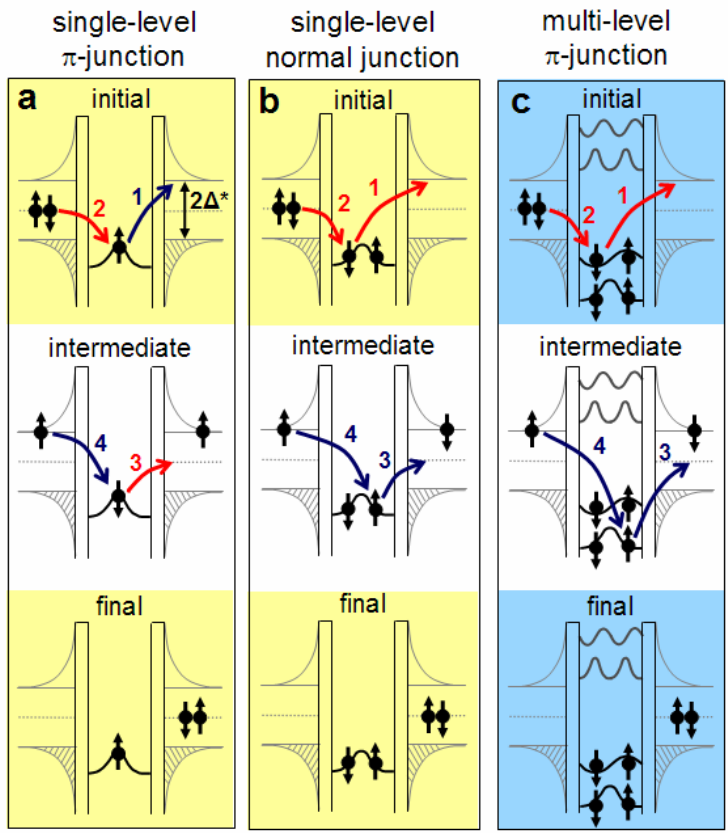

Figure 4

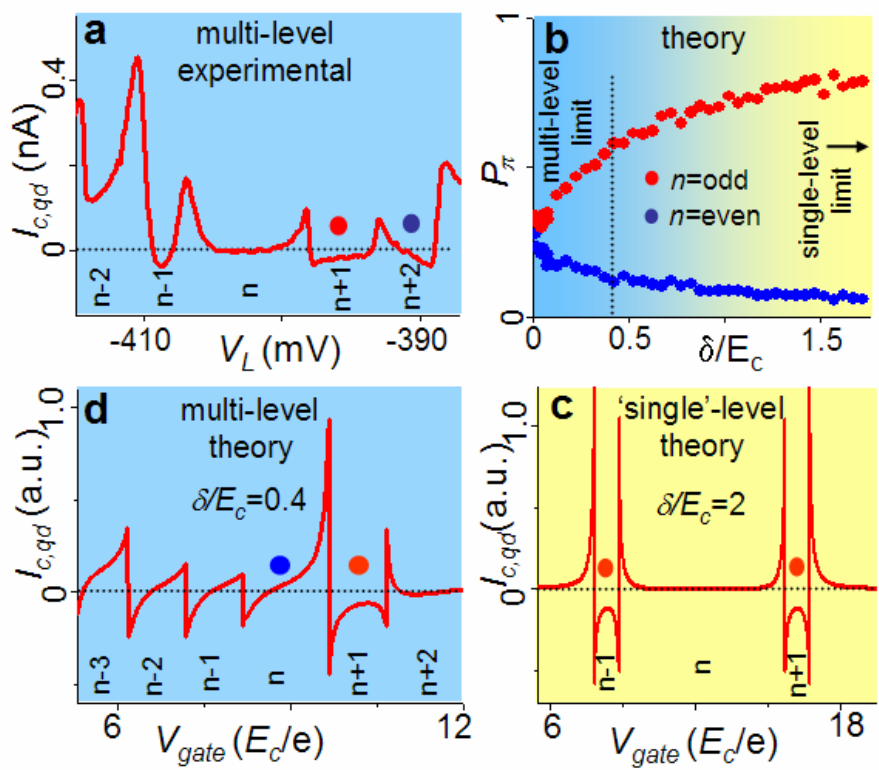

# Supplementary Information

*'Supercurrent reversal in quantum dots'*

## Supplementary methods

*Nanowire growth and device fabrication*

Substrates for the wire growth were prepared by dispersing 20 nm Au colloids on an epi-ready InP(100) substrate. The nanowires were grown epitaxially in the two [111]B directions in the VLS growth mode by the use of a low-pressure (50 mbar) Metal-Organic Vapour-Phase Epitaxy (MOVPE) system (Aixtron 200) (Fig. S1a). Trimethylindium (TMI), phosphine ($PH_3$), and arsine ($AsH_3$) were used as precursors in a total flow of 6.0 l/min where hydrogen ($H_2$) was used as carrier gas. The TMI molar fraction was $2.8 \cdot 10^{-5}$, and the $PH_3$ and $AsH_3$ molar fractions were $1.5 \cdot 10^{-2}$. A $PH_3$ pressure was applied during the pre-anneal (at 550ºC for 10 min) and during the heating of the substrate to the desired growth temperature (420ºC), at which growth was initiated by opening the TMI source. In order to reduce the tapering of the nanowires first an InP segment was grown for 1 minute, followed by the InAs segment, grown for 10 minutes. Although this approach reduces the tapering, the tapering of the InAs nanowires (due to non-catalytic side deposition) is still considerable, as is shown in Fig. S1b. The diameter of the InAs nanowires ranges from ~20 nm at the top of the nanowire to ~70 nm at the base.

After growth, the wires were randomly deposited on a degenerately doped silicon wafer covered with a 250 nm dry thermal oxide. The position of the nanowires was determined using a set of pre-deposited markers. By conventional e-beam lithography the electrodes were defined in a double-layer of PMMA. The superconducting contacts (10nm Ti / 100nm Al) were deposited by e-beam evaporation in a UHV system with a background pressure of $3 \cdot 10^{-8}$ mbar. Before evaporation a 5s BHF-dip was performed in order to reduce the contact resistance. The thin titanium layer ensures a high transparency of the metal/semiconductor interface and the aluminium is used for its superconducting properties. From transport measurements we observed that the superconducting gap ($2\Delta$) of the bi-layer is 200-250µeV. In a second lithography step the local gates were defined and again e-beam deposition was used to deposit 100nm thick aluminium gates. The typical breakdown voltage of the local gates is -2V, which is well below the voltages that we need to locally deplete the InAs nanowires (down to -1V).

## Supplementary data

Here we show additional data for a different device (see Fig. S1c). The general behaviour of this sample is similar to that of the device discussed in the main text. The supercurrent reversal is demonstrated in Fig. S2a, which shows the critical current of the nanowire quantum dot junction, $I_{c,qd}$, as a function of gate voltage, $V_L$ (the critical current of the reference junction, $I_{c,REF}$, is set to 280 pA). The corresponding measurement of differential conductance of the quantum dot, $dI/dV$, as a function of $V$ and $V_L$ is shown in Fig. S2b. A negative supercurrent is observed in two adjacent diamonds. Therefore supercurrent reversal occurs for both odd and even numbers of electrons on the quantum dot. From Fig. S2b we can estimate the ratio of the mean level spacing and the charging energy: $\delta/E_c \approx 0.3$. If we compare this value with the theoretical results presented in Fig. 4b, we indeed expect supercurrent reversal for both odd and even numbers of electrons.

## Supplementary discussion

As explained in the main text two effects determine the sign of supercurrent: (i) permutation of tunnel events, and (ii) parity of wave functions of the levels in the dot. First we discuss both effects separately and then the combination of the two.

(i) Many co-tunnelling events contribute to the supercurrent. Each co-tunnelling event involves 4 elementary tunnel events. The sign of a contribution of a particular co-tunnelling event depends on the sequence of the elementary tunnel events. Each tunnel event is represented by electron creation/annihilation operators, so the sequence of the tunnel events corresponds to a sequence of electron operators. We define operators that create electrons in the right superconducting lead, $c_\sigma^\dagger$, and annihilate electrons in the left superconducting lead, $c_\sigma$, with spin $\sigma$. In order to determine the sign of each contribution we take the sequence of operators and permute operators to achieve the conventional sequence for Cooper pair transport from the left to the right lead: $c_\uparrow^\dagger c_\downarrow^\dagger c_\downarrow c_\uparrow$. Owing to the anti-commutation of electron operators, odd (even) numbers of permutations result in a negative (positive) sign of the contribution.

In general, there are 4!=24 different possible sequences of 4 operators. However, in each concrete situation the number of allowed sequences can be reduced because the levels involved in the co-tunnelling event are either empty or filled.

In the case of a single level filled with one electron, there are 6 sequences allowed. All 6 give a negative contribution to the supercurrent. We demonstrate this for the sequence illustrated in Fig. 3a. The corresponding sequence of operators is given by: $c_\uparrow c_\downarrow^\dagger c_\downarrow c_\uparrow^\dagger$. In order to achieve the conventional sequence for Cooper pair transport we have to perform five permutations. This gives a negative sign:

$$c_\uparrow c_\downarrow^\dagger c_\downarrow c_\uparrow^\dagger = (-1) c_\downarrow^\dagger c_\uparrow c_\downarrow c_\uparrow^\dagger = (-1)^2 c_\downarrow^\dagger c_\downarrow c_\uparrow c_\uparrow^\dagger = (-1)^3 c_\downarrow^\dagger c_\downarrow c_\uparrow^\dagger c_\uparrow =$$
$$(-1)^4 c_\downarrow^\dagger c_\uparrow^\dagger c_\downarrow c_\uparrow = (-1)^5 c_\downarrow^\dagger c_\uparrow^\dagger c_\downarrow c_\uparrow.$$

If the single level is filled with two electrons, the six sequences mentioned before are forbidden by the Pauli exclusion principle. In this situation other sequences are allowed that give a positive contribution. For example, the sequence of operators corresponding to the co-tunnelling event illustrated in Fig. 3b is given by: $c_\uparrow c_\uparrow^\dagger c_\downarrow c_\downarrow^\dagger$. The occurrence of four permutations compared to the conventional sequence of operators for Cooper pair transport results in a positive sign. Therefore the sign of the supercurrent for a single-level dot is determined by the number of electrons on the quantum dot.

In the case of a multi-level quantum dot the above mentioned restrictions on the sequences of operators are relaxed and, in principle, all 24 sequences are possible. As a result, also events can occur that give a negative (positive) contribution to the supercurrent for even (odd) numbers of electrons. To illustrate this we consider the situation when two electrons are transported through different levels for an even number of electrons on the quantum dot (Fig. S3a). In this case the sequence of operators is given by: $c_\uparrow c_\downarrow^\dagger c_\downarrow c_\uparrow^\dagger$. This sequence is identical to the sequence in Fig. 3a, thus resulting in a negative contribution to the supercurrent. By investigating all 24 possible sequences the following general conclusion can be obtained[7]: The contribution to the supercurrent for a co-tunnelling event is negative when one electron is transported through a filled level and the other electron through an empty level. The contribution is positive when both electrons are transported through filled or through empty levels. Note that it is necessary to include dot-operators in order to determine the correct supercurrent sign for all sequences.

(ii) Orbital effects can result in negative supercurrents when two electrons are transported through different levels (or orbitals) of the quantum dot[7]. We illustrate this using Eq. 1 (see next section of Supplementary Information). In this equation, the contribution to the supercurrent of a co-tunnelling event (involving energy levels $i$ and $j$) is proportional to $T_i$ and $T_j$. The sign of $T$ is positive or negative depending on the parity of the corresponding wavefunction. Therefore, when two levels are involved with wavefunctions of opposite parity, the contribution to the supercurrent is negative. An example of such a co-tunnelling event is illustrated in Fig. S3b.

The combined result of effects (i) and (ii) are summarized in Fig. S3c, showing the supercurrent sign for the four possible situations. Note that when both effects are present in one co-tunnelling effect, the contribution to the supercurrent will be positive.

The sign of the supercurrent of the quantum dot is determined by the dominating type of co-tunnelling events (i.e. + or -). As mentioned before, for a single-level quantum dot the supercurrent sign is negative (positive) for odd (even) numbers of electrons on the dot. In the case of a multi-level quantum dot the supercurrent sign is determined by the dominant co-tunnelling events. Co-tunnelling events will have a large contribution to the supercurrent when: 1. the amplitude of the four tunnel couplings is large, and 2. the energy of the intermediate virtual states is small. By numerical evaluation (discussed below) the sign and magnitude of the critical current for a quantum dot with a specific energy spectrum can be calculated.

## Supplementary methods
*Numerical Evaluation of the supercurrent*

To model the quantum dot, we proceed in a conventional way: We introduce a system of discrete spin-degenerate levels with energies $E_i$. In a given charge configuration, these energies are counted from the last level filled. Coulomb interaction is taken into account in addition/extraction energies of the dot. For instance, the energy cost to put an extra electron to the level $i$ reads $E_i + E^+$, to put two electrons to the levels *i* and *j* reads $E_j + E_i + E^{++}$, $E^+, E^{++}$ being the charging energy differences. The charging energy of the state with N electrons reads as usual: $E_{ch} = E_C(N - C_g V_g)^2$.

In a common quantum dot, the actual value of *N* is determined from the minimum of the charging energy and changes in a step-like fashion with increasing gate voltage.

The tunnelling between a discrete level *i* of the dot and a continuous-spectrum state **k** in a lead is generally described by an amplitude $t_{i,\mathbf{k}}$. Due to time-reversibility, all amplitudes can be chosen real. As mentioned, the amplitude of the Cooper pair transfer is contributed by co-tunneling "events", each involving up to two levels (*i,j*) and four elementary tunneling "events". Each co-tunneling event comes with a certain combination of amplitudes: $t_i^L t_i^R t_j^L t_j^R$. It is convenient to introduce products of amplitudes that characterize tunnelling via a certain level: $T_i = t_i^L t_i^R \sqrt{\nu_L \nu_R}$ ; $\nu_L, \nu_R$ being the densities of states on both sides of the contact. We note that $T_i$ can be either positive or negative. Its sign is determined by the *parity* of the corresponding wave-function: positive if the wave-function is of the same sign at both tunnel point contacts and negative otherwise. A compact expression for the Josephson amplitude reads:

$$-E_J = \sum_{i,j,\alpha} T_i T_j \left( A_{ee} \tilde{f}_{i,\alpha} \tilde{f}_{j,-\alpha} + A_{hh} f_{i,\alpha} f_{j,-\alpha} - A_{he} f_{i,\alpha} \tilde{f}_{j,-\alpha} \right);$$

(Eq. 1)

Here, $\alpha$ denotes the spin index and $f_{i,\alpha}$ represents the electron filling factor of a given level ($\tilde{f}_{i,\alpha} \equiv 1 - f_{i,\alpha}$ is the hole filling factor). Since we disregard temperature $f_{i,\alpha}$ can only take two values: 0 and 1. We keep the spin index to treat completely filled and half-filled levels on equal footing: For filled levels $f_{i,\alpha} = f_{i,-\alpha}$, while for half-filled ones $f_{i,\alpha} + f_{i,-\alpha} = 1$. $A_{ee}, A_{hh}, A_{he}$ are positive functions of the two level energies, $\Delta$, and the charging energy. From this expression one can inherit all sign rules previously discussed. For a single half-filled level, *i=j* and only the third term survives resulting in a negative sign. Terms with *i=j* that correspond to either filled or empty levels always provide a positive contribution. For a contribution of a pair of different levels, $i \neq j$, and the sign depends on the sign of $T_i T_j$, that is, on relative parity of the corresponding wave-functions. If both wave-functions are odd or even, the contribution is positive if both levels are filled or empty, and negative otherwise. If one wave-function is odd and another one is even, the situation is opposite: the sign is negative if both levels are filled or empty, and positive otherwise. The concrete expressions for $A_{ee}, A_{hh}, A_{he}$ read:

$$A_{ee}(E_i, E_j) = \int d\varepsilon_1 d\varepsilon_2 \nu(\varepsilon_1) \nu(\varepsilon_2) \left( \frac{1}{(\varepsilon_1 + E_i + E^+)(E_i + E_j + E^{++})(\varepsilon_2 + E_j + E^+)} + \frac{1}{(\varepsilon_1 + E_i + E^+)(E_i + E_j + E^{++})(\varepsilon_2 + E_i + E^+)} + \frac{1}{(\varepsilon_1 + E_i + E^+)(\varepsilon_1 + \varepsilon_2)(\varepsilon_2 + E_j + E^+)} \right)$$

$$A_{hh}(E_i, E_j) = \int d\varepsilon_1 d\varepsilon_2 \nu(\varepsilon_1) \nu(\varepsilon_2) \left( \frac{1}{(\varepsilon_1 - E_i + E^-)(-E_i - E_j + E^{--})(\varepsilon_2 - E_j + E^-)} + \frac{1}{(\varepsilon_1 - E_i + E^-)(-E_i - E_j + E^{--})(\varepsilon_2 - E_i + E^-)} + \frac{1}{(\varepsilon_1 - E_i + E^-)(\varepsilon_1 + \varepsilon_2)(\varepsilon_2 - E_j + E^-)} \right)$$

$$A_{he}(E_i, E_j) = \int d\varepsilon_1 d\varepsilon_2 \nu(\varepsilon_1)\nu(\varepsilon_2) \begin{pmatrix} \dfrac{1}{(\varepsilon_1 - E_i + E^-)(-E_i + E_j + \varepsilon_1 + \varepsilon_2)(\varepsilon_2 - E_i + E^-)} + \\ \dfrac{1}{(\varepsilon_2 + E_j + E^-)(-E_i + E_j + \varepsilon_1 + \varepsilon_2)(\varepsilon_2 + E_j + E^-)} + \\ \dfrac{2}{(\varepsilon_1 + E_j + E^+)(\varepsilon_1 + \varepsilon_2)(\varepsilon_2 - E_i + E^-)} + \\ \dfrac{2}{(\varepsilon_1 - E_i + E^-)(\varepsilon_1 + \varepsilon_2 - E_i + E_j)(\varepsilon_2 + E_j + E^+)} \end{pmatrix}$$

In all cases, they are obtained by integration of three energy denominators corresponding to the possible virtual states over energies of virtual quasiparticles. $\nu(\varepsilon_1)$ in the above expressions presents the BCS factor of the superconducting density of states, $\nu(\varepsilon) \equiv \Theta(\Delta^2 - \varepsilon^2)|\varepsilon|/\sqrt{\varepsilon^2 - \Delta^2}$. Two-dimensional integration required for the numerical evaluation of $A$ slowed down the simulations considerably.

The concrete numerical calculations have been performed in two ways. Firstly, we take a randomly chosen realisation of $T_i, E_i$ and calculate the current as a function of gate voltage in a wide interval of gate voltage. Typically, we took about hundred levels and the interval of gate voltage where the number of electrons varied from 50 to 70. The ratio of charging energy and $\Delta$ was fixed to the experimental value while the ratio of level spacing and charging energy has been varied in a wide range. The goal of this simulation was to compare typical patterns in critical current-gate voltage dependence with those observed experimentally. The critical current typically showed pronounced peaks at the gate voltages where the number of particles changes (edges of diamonds). Peaks of either sign have been obtained. Randomly distributed $T_i$ result in strong (by order of magnitude) variations of the current from diamond to diamond. The supercurrent changes sign within a diamond as well as at the edges. If the average spacing was smaller than the charging energy, the patterns and the magnitudes of the current exhibit a relatively strong correlation in neighbouring diamonds. This indicates that the current in this case is contributed by many levels, those are essentially the same in neighbouring diamonds. The situation is opposite for a large level spacing where the level closest to the Fermi energy clearly dominates the supercurrent. This, as discussed, gives a positive (negative) supercurrent for even (odd) numbers of electrons.

Secondly, we use the same simulation scheme to quantify the probability of a negative/positive sign of the supercurrent. In this case, we fix the gate voltage to the middle of an even (odd) diamond and evaluate the current for a big set of random realizations of $T_i$ (typically, 10.000 realizations). So we can get the probability with 1% accuracy.

*Note 1:* It is important to recognize an important detail specific for a dot connected to superconducting leads: its charging state is *bistable* at zero temperature near the values of gate voltage corresponding to the charge of *N*. This is a consequence of the fact that one has to create a quasiparticle in order to put a charge into a superconductor, this costs extra energy $\Delta$. In our simulations, we disregarded the bistability assuming that the dot is always in the ground state. This should correspond to the experimental situation where the supercurrent has been measured at small but finite voltage. This voltage, although small, may generate the quasiparticles required for a fast relaxation to the ground state.

*Note 2:* We *disregard* the dependence of the amplitude on **k**. This corresponds to the important physical assumption of a *point-like* tunnelling contact where most tunnelling processes take place within the same transport channel that has the highest transparency. While this assumption frequently fails for natural oxide tunnel barriers in metallic systems, it is well-justified and proven for electrostatically formed tunnel barriers in semiconductor quantum dots where the potential profile is smooth at the scale of the electron wave-length. Since the junctions in our experiment are formed electrostatically, we assume that the tunnel contacts are point-like.

# Figure S1

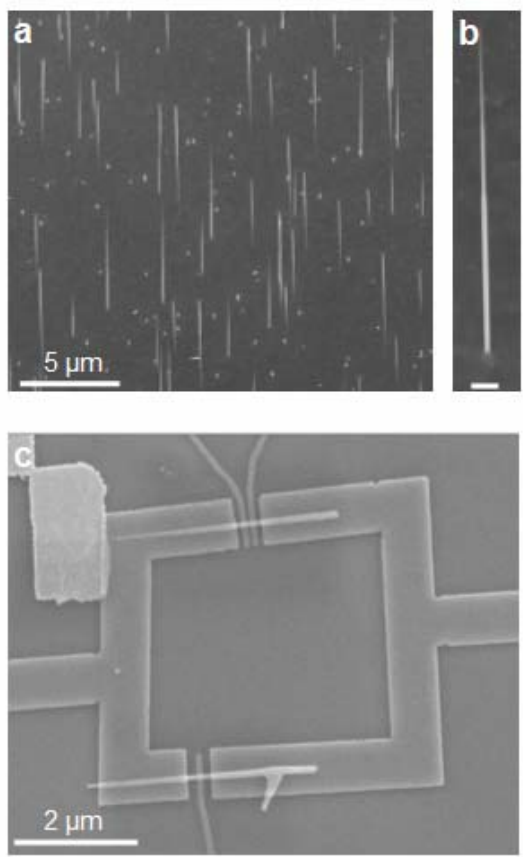

**Figure S1.** Scanning Electron Microscopy (SEM) images. **a**, SEM-image of InAs nanowires epitaxially grown on an InP(100) substrate. **b**, SEM-image of an InAs nanowire (scalebar is 300 nm). **c**, SEM-image of the second nanowire device. An alignment marker is visible in the upper left part of the image.

Figure S2

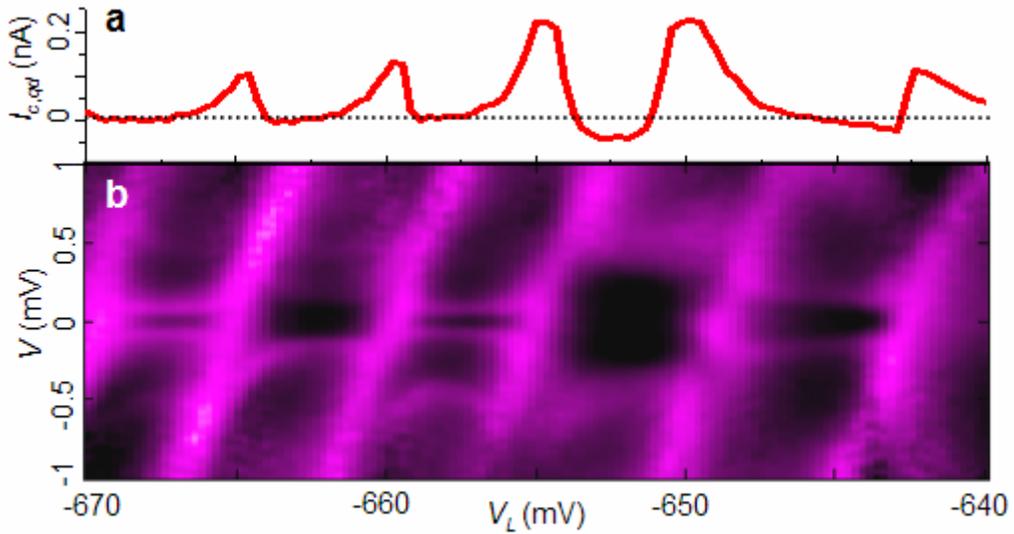

**Figure S2.** Supercurrent reversal in the second device. **a,** Plot of the critical current of the quantum dot, $I_{c,qd}$, as a function of gate voltage, $V_L$, for the same gate voltage region as in (b). **b,** Colour scale plot of differential conductance, $dI/dV$, as a function of bias voltage, $V$, and gate voltage, $V_L$ ($dI/dV$ increases from black to purple). Two adjacent charge states exhibit $\pi$-behaviour.

# Figure S3

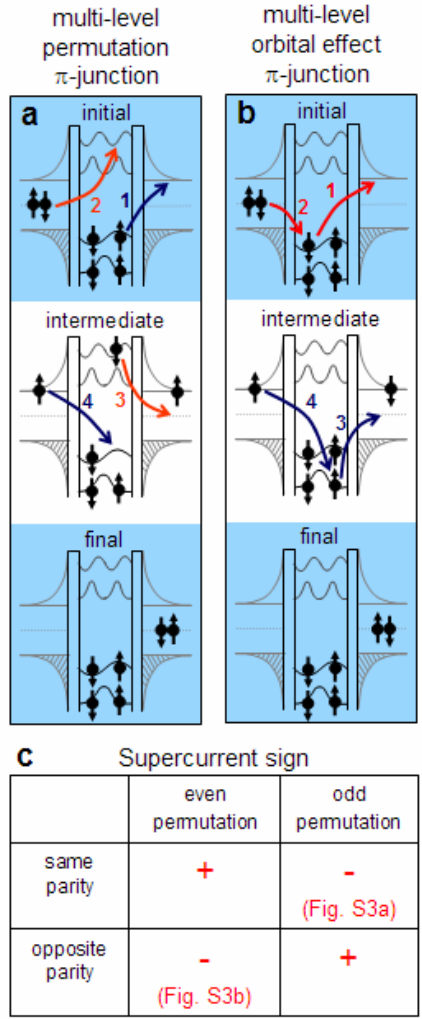

**Figure S3.** Energy diagrams illustrating transport through a mulit-level quantum dot. **a**, Transport occurs through a filled and an empty energy level. The corresponding sequence of operators involves five permutations resulting in a negative supercurrent. **b**, Transport occurs through two filled energy levels with opposite parity of the corresponding wavefunctions. This results in a negative contribution to the supercurrent. **c**, Sign of supercurrent contribution for the four different types of co-tunnelling events. Only if one of the above effects occurs, the contribution to the supercurrent is negative.